\documentclass[a4paper,twocolumn,aps,superscriptaddress,showpacs,showkeys]{revtex4-1}
\usepackage[latin9]{inputenc}
\usepackage{amsmath}
\usepackage{amssymb}
\usepackage{graphicx}

\usepackage[colorlinks=true, linkcolor=red, citecolor=blue, urlcolor=blue]{hyperref}
\usepackage[usenames,dvipsnames]{color}
\usepackage{bm}
\usepackage{bbold}
\usepackage{braket}
\usepackage{float}
\usepackage{comment}
\usepackage{dcolumn}

\definecolor{mygreen}{rgb}{0,0.5,0}
\definecolor{myblue}{rgb}{0,0,0.75}
\definecolor{mymagenta}{cmyk}{0,1,0,0.12}

\newcommand{\minus}{
  \setbox0=\hbox{-}
  \vcenter{
    \hrule width\wd0 height \the\fontdimen8\textfont3
  }%
}

\def\po{\tfrac{1}{2}}
\def\mo{\minus\tfrac{1}{2}}

\def\inner(#1,#2,#3,#4,#5,#6){\ensuremath\left(\begin{array}{ccc} #1 & #2 & #3 \\ #4 & #5 & #6 \end{array}\right)}

\def\innerv(#1,#2,#3,#4,#5,#6){\ensuremath\left\{\begin{array}{ccc} #1 & #2 & #3 \\ #4 & #5 & #6 \end{array}\right\}}

\begin{document}

\title{Simulating Quantum Spin Models using Rydberg-Excited Atomic Ensembles in Magnetic Microtrap Arrays}

\author{Shannon Whitlock}
\email{whitlock@uni-heidelberg.de}
\affiliation{Physikalisches Institut, Universität Heidelberg, Im Neuenheimer Feld 226, 
69120 Heidelberg, Germany}

\author{Alexander W. Glaetzle}
\affiliation{Institute for Quantum Optics and Quantum Information of the Austrian
Academy of Sciences, A-6020 Innsbruck, Austria}
\affiliation{Institute for Theoretical Physics, University of Innsbruck, A-6020
Innsbruck, Austria}

\author{Peter Hannaford}
\email{phannaford@swin.edu.au}
\affiliation{Centre for Quantum and Optical Science, Swinburne University of Technology,
Melbourne, Victoria 3122, Australia}

\date{\today}
\begin{abstract}
We propose a scheme to simulate lattice spin models based on strong and long-range interacting Rydberg atoms stored in a large-spacing array of magnetic microtraps. Each spin is encoded in a collective spin state involving a single $nP$ Rydberg atom excited from an ensemble of ground-state alkali atoms prepared via Rydberg blockade. After the excitation laser is switched off the Rydberg spin states on neighbouring lattice sites interact via general isotropic or anisotropic spin-spin interactions. To read out the collective spin states we propose a single Rydberg atom triggered avalanche scheme in which the presence of a single Rydberg atom conditionally transfers a large number of ground-state atoms in the trap to an untrapped state which can be readily detected by site-resolved absorption imaging. Such a quantum simulator should allow the study of quantum spin systems in almost arbitrary two-dimensional configurations. This paves the way towards engineering exotic spin models, such as spin models based on triangular-symmetry lattices which can give rise to frustrated-spin magnetism.
\end{abstract}



\maketitle

\section{Introduction}
Periodic arrays of quantum spins coupled through magnetic interactions represent an archetypal model system in quantum many-body physics, non-equilibrium physics, statistical physics and condensed matter physics, with potential implications ranging from quantum magnetism to quantum information science, spintronics and high-temperature superconductivity~\cite{Sachdev1999,Lewenstein2007,Bloch2008}.  Apart from a few special cases, such models are generally computationally intractable due to extreme complexity arising from quantum entanglement between the spins. Furthermore, experimental studies on solid-state spin systems are often restricted by uncontrolled disorder and random couplings to the environment as well as limited control over system parameters. 

\begin{figure}[tb]
\centering
\includegraphics[width= .9 \columnwidth]{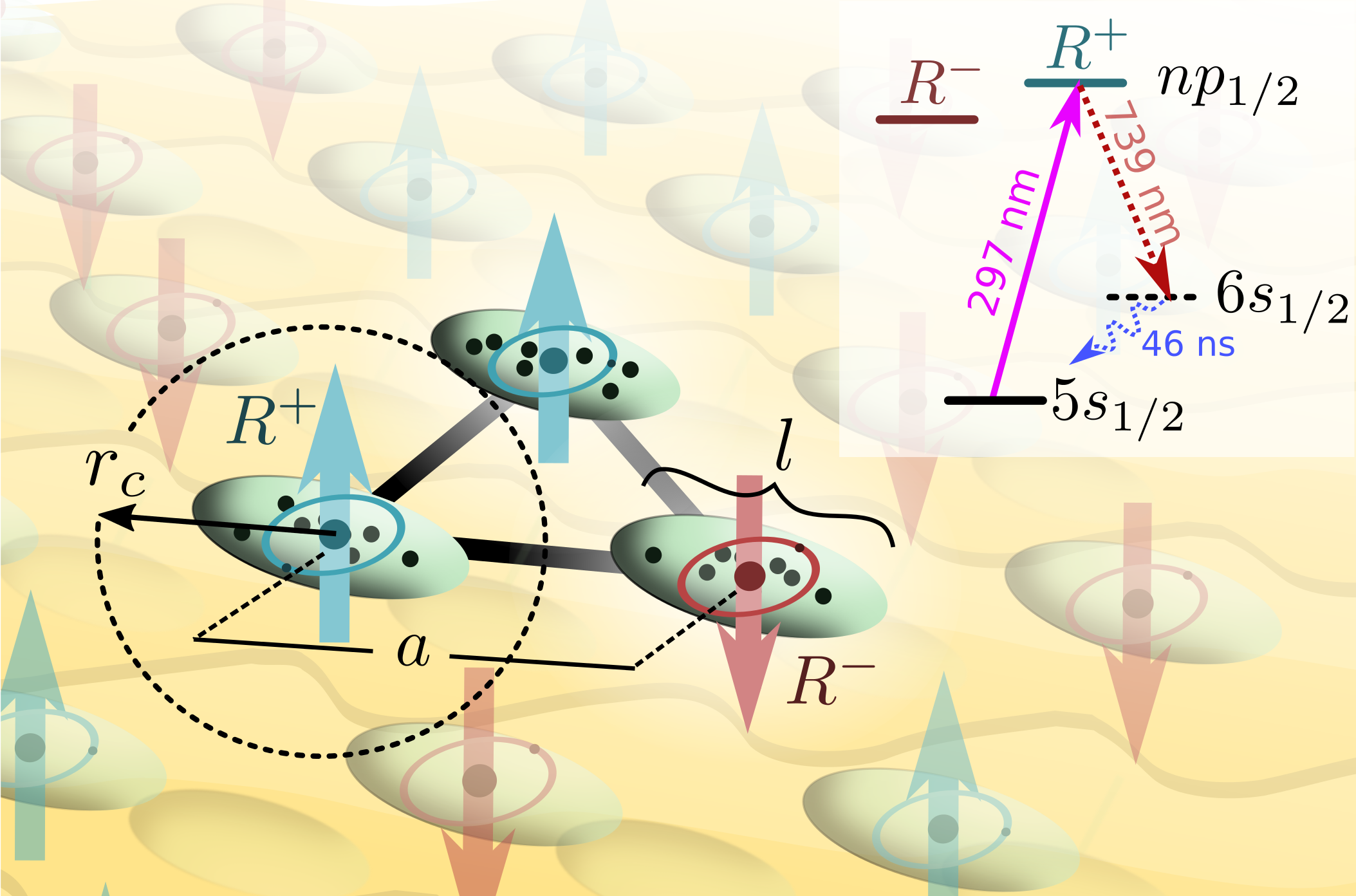}
\caption{\small{Proposed experimental setup for simulating quantum spin models in atomic ensembles confined in a microtrap array. Two spin states $\mid\uparrow\rangle$ and $\mid\downarrow\rangle$ are encoded in a single excitation to the Rydberg state $\ket{R^+}$ or $\ket{R^-}$, shared amongst all atoms in an ensemble. The spatial extent of each ensemble is $\ell$, the on-site blockade radius is $r_c$ and the lattice spacing is $a$, with $a\gtrsim r_c\gg\ell$. The inset shows the internal level structure of a single atom with the states involved in the detection processes marked with dashed lines. General spin-spin interactions occur via long-range and anisotropic van der Waals interactions between the $\ket{R^+}$ and $\ket{R^-}$ states.}}
\label{fig1}
\end{figure}

There is currently a growing interest in utilising ultracold atoms stored in periodic lattices to simulate many-body and condensed matter systems (e.g.,~\cite{Jaksch2005,Lewenstein2007, Bloch2008,Simon2011,Hart2015,Zeiher2016}).  Ultracold atoms trapped in periodic lattices allow precise control over system parameters, such as the inter-particle interaction, lattice geometry and disorder, and, in principle, provide an ideal platform to achieve almost perfect realisations of a variety of lattice spin models~\cite{Jaksch2005,Lewenstein2007, Bloch2008, Simon2011,Hart2015,Zeiher2016}. Most proposals to date have been based on tunnelling and on-site interactions in optical lattices, in which the magnetic interaction energy scales are set by a super-exchange interaction whose strength scales as $J\sim t^2/U$~\cite{Lewenstein2007} (with $t$ the tunnelling rate and $U$ the on-site interaction energy). The $t^2$-dependence results in extremely small magnetic energy scales, of the order of only a few tens of hertz, corresponding to just a few nanokelvin. Thus, with spin models based on tunnelling dynamics, a major experimental challenge is the requirement of extremely low temperatures, close to or beyond the limits of current state-of-the-art atom cooling techniques~\cite{Medley2011,Hart2015}. Possible ways to circumvent this very low-temperature requirement have been proposed, including the use of ultracold polar molecules~\cite{Buchler2007} and Rydberg-dressed ground-state atoms~\cite{Glaetzle2015,Zeiher2016} featuring widely tunable long-range interactions which can be effective over distances much larger than typical optical lattice periods.

In this paper we propose the use of long-range interacting Rydberg atoms prepared in a large-spacing (several $\mu$m) lattice of magnetic microtraps~\cite{Ghanbari2006,Singh2008,Whitlock2009,Schmied2010,Jose2014,Surendran2015,Herrera2015} to simulate lattice spin models. This scheme is similar to earlier proposals to create Rydberg quantum gates in mesoscopic ensembles in the context of quantum information science~\cite{Mueller2009,Weimer2011,Nyman2011,Leung2011}. Each spin is encoded in a collective spin state involving a single rubidium $nP$ Rydberg atom in an ensemble of ground-state Rb atoms prepared via Rydberg blockade~\cite{Saffman2010} (Fig.~\ref{fig1}). The use of atomic ensembles avoids the problem of exact single-atom filling of lattice sites and single-atom detection, which is a requirement for schemes based on Rydberg-dressed ground-state atoms~\cite{Glaetzle2015,Zeiher2016}, and also helps greatly with the initialisation and readout of individual Rydberg spin states. The long-range and widely tunable interactions between Rydberg atoms combined with a large-spacing between the interacting spins readily facilitates site-resolved detection using standard optical imaging techniques. {The finite orbital angular momentum of Rydberg $p$-states allows for Heisenberg-like and more general spin-spin interactions which can be isotropic or anisotropic and can extend beyond nearest-neighbours~\cite{Glaetzle2015,Bijnen2015}. A second advantage of using Rydberg $p$-states is that they can be excited directly from the ground state without the complication of an intermediate state. }The timescales associated with atomic motion ($\sim$ ms) or lifetimes of high $nP$ Rydberg states ($>50~\mu$s)~\cite{Beterov2009} are long compared to the timescales associated with strong Rydberg-Rydberg interactions ($\sim$1~$\mu$s), which enables investigation of non-equilibrium spin dynamics on both short and long times, including, for example, the build-up of spin-spin correlations following a sudden quench of the system parameters.

\section{Simulating Quantum Spin Models}
As a concrete experimental platform we consider an array of magnetic microtraps created by patterned magnetic films on an atom chip~\cite{Singh2008,Whitlock2009,Schmied2010,Jose2014,Leung2014,Surendran2015,Herrera2015}. A general algorithm has been developed to design the required magnetic patterns, enabling microtrap arrays to be produced with nearly-arbitrary 2D symmetries and orientations of the magnetic field at the bottom of each microtrap, without restrictions imposed by optical fields~\cite{Schmied2010}. Lattices of magnetic microtraps with triangular and square symmetry with a period of 10 $\mu$m have already been realised and loaded with small atomic ensembles, each consisting of a few hundred atoms~\cite{Leung2011}. 

For our implementation we assume each site $\alpha$ contains an ensemble of $N_\alpha$ rubidium atoms confined to a characteristic size $\ell$ and different sites are separated by the lattice period $a$, with $\ell \ll a$. We consider the following excitation, interaction and detection sequence:

\vskip 5pt
\noindent {\it 1. Initialisation}: Each lattice site is prepared with precisely one Rydberg excitation, e.g., using a single-photon laser excitation at 297~nm tuned to the $|g\rangle\rightarrow |R^+\rangle\equiv |nP_{1/2},m_j=+1/2\rangle$ transition. Assuming Poissonian statistics with a mean number of atoms $\bar N=10$ the probability to load zero atoms in a given site is $<10^{-4}$; therefore we can expect large filling factors. To restrict the system to a single excitation on each site we propose to use the Rydberg blockade effect which strongly suppresses the probability to excite more than one atom in the ensemble~\cite{Saffman2010}. Small amounts of controlled disorder may be introduced to the resulting spin models, either through the presence of empty sites or the random positions of the Rydberg excitations within each cloud that modifies the nearest neighbour spin-spin couplings. Numerical simulations of the initialisation scheme including anticipated experimental limitations on the achievable filling factors is discussed in Sec.~\ref{sec:initialisation}.

\vskip 5pt
\noindent {\it 2. Interaction time}: Following initialisation, the excitation laser is switched off and Rydberg excitations on neighbouring lattice sites can interact as a consequence of their giant electric dipole moments (typically several kilodebye). At large separations and away from F\"orster resonances, Rydberg-Rydberg interactions can be treated perturbatively leading to van der Waals (vdW) interactions which scale as $n^{11}$~\cite{Saffman2010}, {with $n$ the principal quantum number}. We identify two collective spin states for a single site
\begin{equation}
\begin{split}
&\mid\uparrow\rangle=\frac{1}{\sqrt{N}}\sum_j|g_1,\ldots,g_{j-1},R_j^+,g_{j+1},\ldots,g_N\rangle,\\
&\mid\downarrow\rangle=\frac{1}{\sqrt{N}}\sum_j|g_1,\ldots,g_{j-1},R_j^-,g_{j+1},\ldots,g_N\rangle,
\label{eq:spins}
\end{split}
\end{equation}
(see Fig.~\ref{fig1}), where $|R^\pm\rangle$ denotes the $|nP_{1/2},m_j=\pm 1/2\rangle$ Rydberg states. These collective spin states are coherent superpositions with the single Rydberg excitation shared amongst all atoms in the ensemble~\cite{Saffman2010}. This configuration will allow complex spin-spin interactions including Heisenberg-like and more general spin-spin interactions, which can be isotropic or anisotropic as described in Sec.~\ref{sec:spin_interactions}. Additionally, the two collective spin states can be coupled using radiofrequency transitions to realise spin models with effective magnetic fields. 

\vskip 5pt
\noindent {\it 3. Readout}: To read out the collective spin state one needs to be able to detect the presence of a single Rydberg atom in a given spin state in the atomic ensemble with high fidelity. Here, the use of atomic ensembles is a significant advantage. We propose to use a single-Rydberg atom triggered ionisation `avalanche' scheme, similar to recent observations~\cite{Vincent2013,Schempp2014,Simonelli2016}, in which the presence of the single Rydberg atom conditionally transfers a large number of ground-state atoms in the trap to an untrapped state which can then be detected by standard site-resolved absorption imaging, as described in Sec.~\ref{sec:readout}. 

\vskip 5pt
In the following we identify some general criteria for this system to be suitable for the quantum simulation of spin models. First, the typical rate associated with spin-spin interactions between neighbouring sites must greatly exceed the decoherence rate predominantly given by the Rydberg state decay rate $\Gamma$. Second, to prevent evolution of the quantum spin system during the initialisation phase we additionally require that the Rydberg excitation bandwidth exceeds the spin-spin coupling rate between neighbouring ensembles. Finally, we require that the interactions between atoms within each ensemble far exceed the excitation bandwidth to ensure good conditions for the Rydberg blockade. Combining these constraints we can define the following criteria:
\begin{equation}\label{eq:criteria}
\frac{|C_6|}{\ell^6} \gg \sqrt{\bar N}\Omega \gg \frac{|C_6|}{a^6} \gg \Gamma.
\end{equation}
These citeria can be met for typical conditions in a large-spacing magnetic lattice~\cite{Surendran2015}. To illustrate this we take the $\ket{36P_{1/2}}$ state of $^{87}$Rb and assume a lattice with a period $a\approx 2.5\,\mu$m, trap size $\ell\approx 2\sigma = 0.8\,\mu$m and mean number of atoms $\bar N=10$. The Rydberg state decay rate, including decay by spontaneous emission and blackbody {radiation at $T=300$~K, is $\Gamma=2\pi\times(4.2\,\mathrm{kHz}$})~\cite{Beterov2009}. This is much smaller than the spin-spin coupling between neighbouring states $C_6/a^6\approx -2\pi\times(400\,\mathrm{kHz})$. The experimentally achievable Rabi frequency with collective enhancement using commercially available laser sources at 297~nm is $\sqrt{\bar N} \Omega\approx 2\pi\times(3\,\mathrm{MHz})$. The intrasite interaction strength $C_6/\ell^6\gtrsim -2\pi\times(360\,\mathrm{MHz})$. Thus, each of the criteria in Eq.~\eqref{eq:criteria} is satisfied by approximately an order of magnitude or more. A realistic cloud geometry, including the relevant length scales for the $n=36$ state, is shown in Fig.~\ref{fig:geometry}.

\section{Initialisation of Collective Spin States}\label{sec:initialisation}

To initialise the spin lattice we propose to use collectively enhanced atom-light coupling in each microtrap to drive Rabi oscillations between the ground state and a state involving a single Rydberg excitation. Complete population inversion can be realised  by interrupting the dynamics after a fixed duration corresponding to a Rabi $\pi$-pulse. Recent experiments have demonstrated the preparation of atomic ensemble qubits in this way with an efficiency of $\langle p_1\rangle=0.62$~\cite{Ebert2014}. However, it is still unclear how much further this can be increased taking into account realistic experimental conditions.

To find conditions which optimise the probability to end the excitation sequence with precisely one Rydberg excitation in each ensemble, we perform numerical simulations of the Rabi dynamics for a small ensemble of atoms resonantly driven to the Rydberg state. Three main mechanisms are assumed which limit the achievable state preparation efficiency: (1) Poissonian atom number fluctuations in each of the magnetic lattice sites due to the stochastic loading process which leads to some disorder in the collective Rabi frequency. (2) Imperfect blockade due to the finite size of the ensemble and the anisotropic character of the $nP$ state van der Waals interactions and (3) short-range physics associated with Rydberg-molecular states. We treat these as independent effects which allows the identification of the dominant limits in an experiment.

\begin{figure}[tb]
\centering
\includegraphics[width=.6\columnwidth]{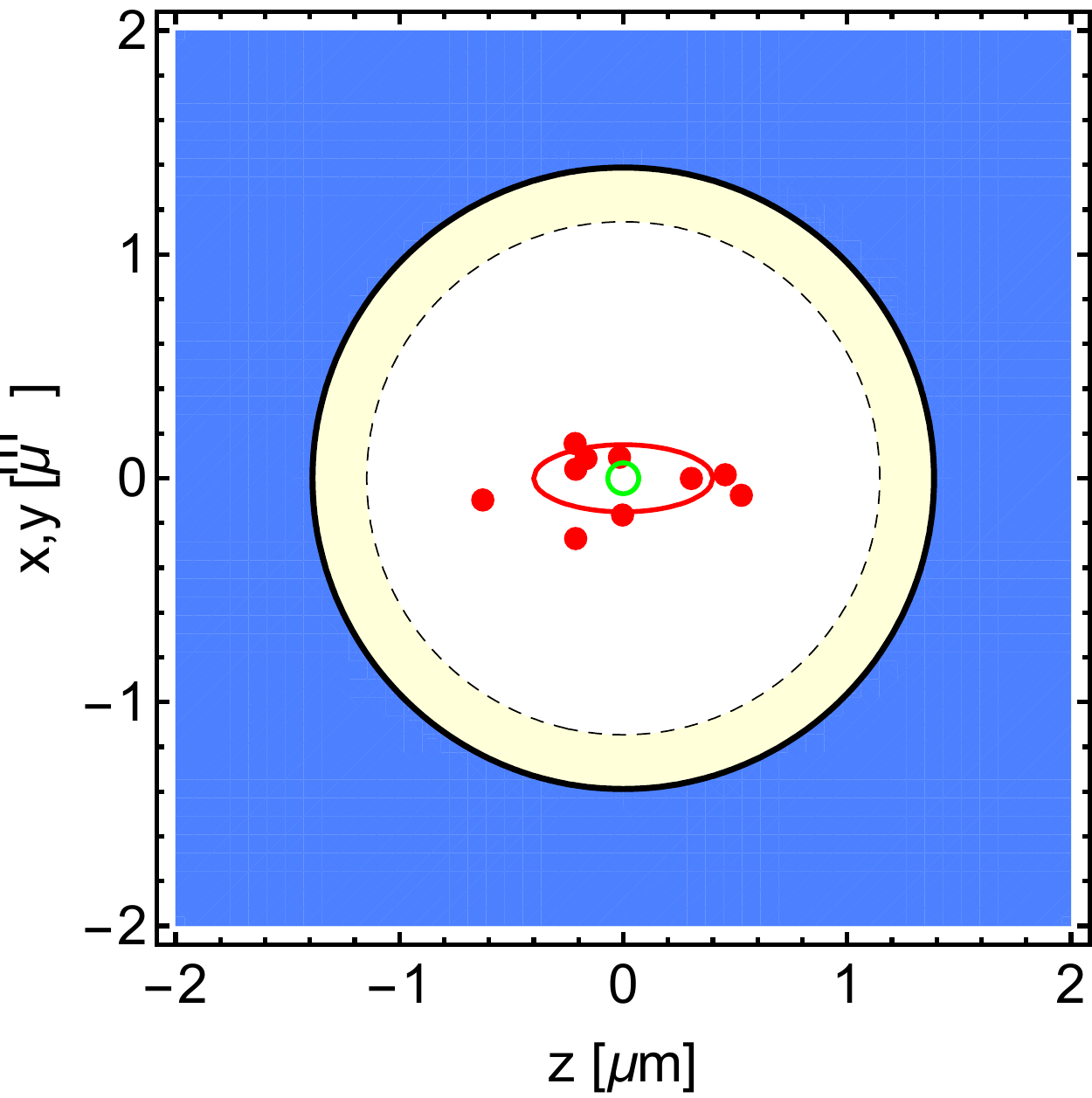}
\caption{\small{Simulated geometry showing an example atomic distribution for a single lattice site and the relevant length scales for $n=36$. The red dots represent $N=10$ individual atoms while the red ellipse represents the $1\sigma$ cloud radii. The outermost contour (solid black line) shows the blockade radius defined as the position where the eigenvalue of $V(r,\vartheta)$ with the smallest magnitude equals the excitation bandwidth given by the Rabi frequency $\Omega$. The inner red dashed contour shows the reduced blockade radius corresponding to the collectively enhanced Rabi frequency $\sqrt{N}\Omega$. The small green circle represents the size of the Rydberg electron wavefunction ($n^2$ in atomic units).}}
\label{fig:geometry}
\end{figure}

\begin{figure}[tb]
\centering
\includegraphics[width=  .9\columnwidth]{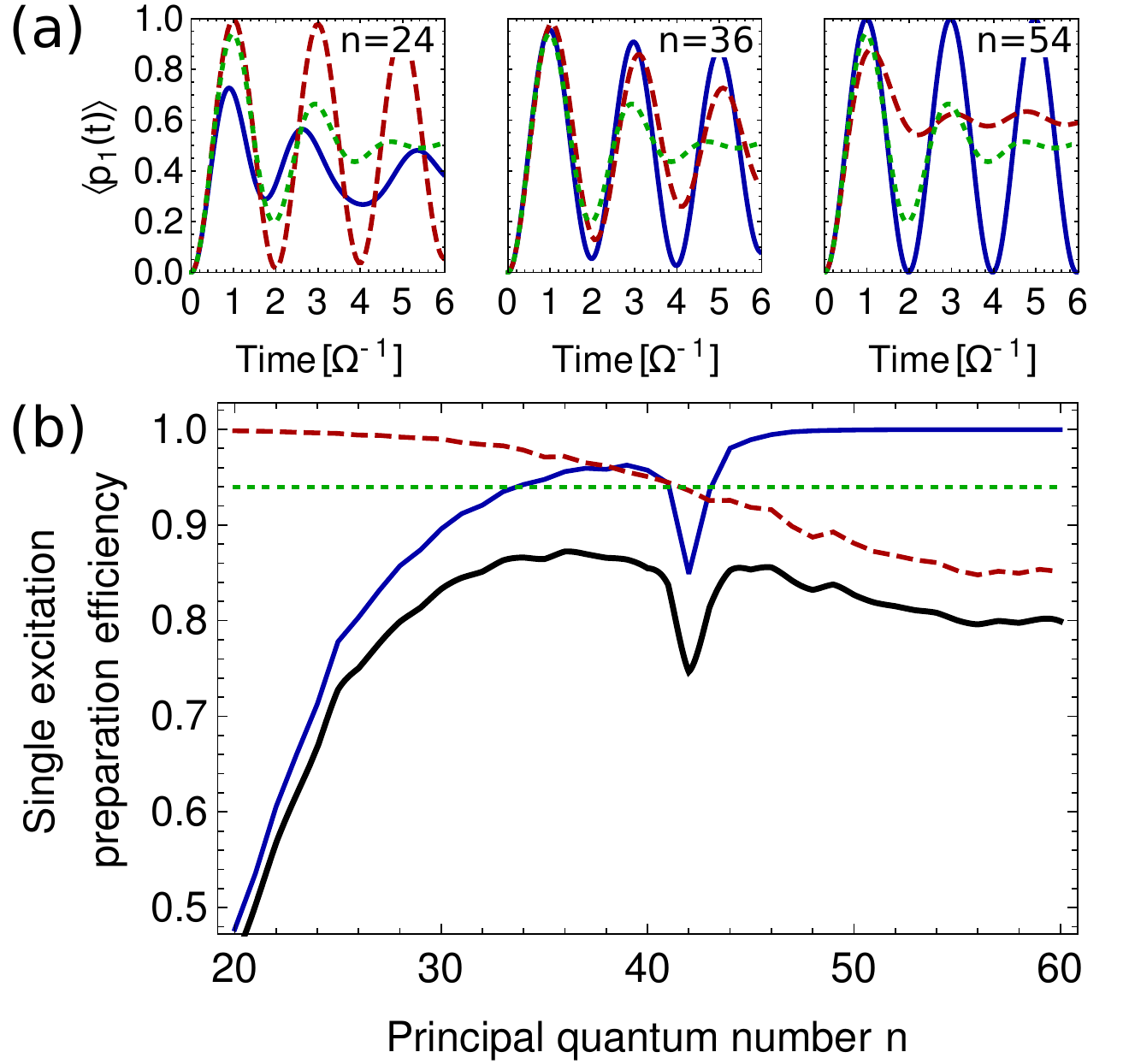}
\caption{\small{Simulated single-excitation preparation efficiencies for an ensemble of $\bar N=10$ atoms in a microtrap with radii $\sigma_x=\sigma_y=0.15\,\mu m,\sigma_z=0.4\,\mu m$. The different lines show the effects of Poissonian atom number fluctuations (dotted green), imperfect blockade (solid blue) and short-range physics due to Rydberg electron-atom scattering (dashed red). (a) Calculated Rabi oscillation curves averaged over 1000 random atomic distributions for three different principal quantum numbers: $n=24$, $n=36$ and $n=54$ (from left to right). (b) Single-excitation preparation efficiency as a function of the principal quantum number $n$. The black line shows the product of the three processes indicating an optimum around $n=36$ and a combined single-atom preparation efficiency of $\approx 0.87$.}}
\label{fig4}
\end{figure}

Figure~\ref{fig4}(b) shows the calculated single excitation preparation efficiency $\langle p_1\rangle$ in the microtrap containing an average of $\bar N=10$ atoms randomly distributed according to an elongated 3D Gaussian distribution with one-sigma radii $\sigma_z=0.4\,\mu$m and $\sigma_{x,y}=0.15\,\mu$m~(as shown in Fig.~\ref{fig:geometry}) and averaged over 1000 runs. We assume a resonant single-atom laser coupling with Rabi frequency $\Omega/2\pi= 1$~MHz. The magnetic field is taken to be oriented along $z$ (trap long axis) and for the initial Rydberg state we use $|nP_{1/2},m_j=1/2\rangle$, where $n$ is varied in order to obtain the best single excitation preparation efficiency.

The first set of simulations quantifies the role of Poissonian atom number fluctuations assuming perfect Rydberg blockade [horizontal dotted line in Fig.~\ref{fig4}(b)]. In this case the dynamics can be reduced to an effective two-level problem with $\sqrt{N}$ enhanced atom-light coupling. The probability to end the sequence with a single excitation oscillates according to $p_1(t)=1-\cos^2(\sqrt{N}\Omega t/2)$, see Fig.~\ref{fig4}(a). Assuming $N$ is sampled from a Poisson distribution with mean $\bar N$, we calculate the excited-state probability after a time $\tau = \pi/(\sqrt{\bar N}\Omega)$ (corresponding to a $\pi$-pulse for the collective Rabi oscillations) and average over $N$. For $\bar N=10$ this gives $\langle p_1(\tau)\rangle=0.94$ independent of the principal quantum number $n$ (assuming perfect blockade). A simple approximation for the average single-excitation preparation efficiency in the limit $\bar N \gg 1$ can be obtained by expanding $p_1(\tau)$ to second order in $N-\bar N$ yielding $\langle p_1(\tau)\rangle\approx 1-\pi^2/(16\bar N)$. 

The second set of simulations concerns the role of the finite size of the ensemble and the imperfect blockade on the efficiency for preparing a single Rydberg excitation [blue solid line in Fig.~\ref{fig4}(b)]. Here, we assume each ensemble contains a precise number of atoms randomly distributed within the Gaussian shaped cloud. Interactions between different Rydberg pair states are well approximated by anisotropic vdW interactions~(Fig.~\ref{fig:geometry}) with coefficients calculated for the $nP_{1/2}$ states of $^{87}$Rb for each value of $n$~(see Appendix~\ref{app:vdW}). However, within each microtrap the interactions are much larger than the Zeeman splitting between the two relevant spin states $\ket{R^\pm}$~\cite{walker2008consequences}. Therefore, we diagonalise the $4\times 4$ interaction matrix and take the eigenvalue with the smallest magnitude which results in an isotropic interaction potential and provides a lower bound to the interaction strength between pairs. The state of the system, including the atom-light coupling, is evolved according to the $N$-atom Schr\"odinger equation as a function of time using a reduced Hilbert space truncated at a maximum of three Rydberg excitations in the ensemble. For principal quantum numbers $n\lesssim 35$ the vdW interactions are not sufficient to completely prevent double excitations in the ensemble, leading to more complicated multilevel Rabi dynamics which after ensemble averaging leads to a damping of the single excitation preparation efficiency. Around $n=42$ a dip in the single-excitation preparation efficiency is observed which is attributed to a sign change of the vdW potential for certain orientations (c.f. Fig.~9 of Ref.~\cite{Vermersch2015}) which reduces the effective blockade radius. 

The third set of simulations takes into account the level shifts induced by the interaction between a Rydberg electron and the surrounding ensemble atoms acting as perturbers in the $nP_{1/2}+5S_{1/2}$ potential [red dashed line in Fig.~\ref{fig4}(b)]. We are not concerned with Rydberg-atom-pair-states (macrodimers) since for the small pair distances within the microtraps there are relatively few pair-states with significant $nP$ character which can be coupled from the ground state and these molecular potential curves are extremely steep leading to a small Frank-Condon factor (see Appendix~\ref{app:shortrange}). Following Ref.~\cite{Schlagmuller2016}, we calculate the energy shift for a given configuration using the Fermi pseudopotential approach and the measured value of the electron-rubidium s-wave triplet scattering length $a_s=\nobreak-15.7\,a_0$~\cite{Bottcher2016}. Higher partial wave scattering is not expected to have a dramatic effect on the short-time Rabi dynamics for the considered densities. The energy shift of the ensemble for a given configuration of atoms {is} given by
\begin{equation}
\Delta_R=\frac{2\pi \hbar a_s}{m_e}\sum_{i}\rho_e(\vec r_i-\vec r_R). 
\end{equation}
where $\vec{r}_i$ and $\vec r_R$ denote to the position of the ground state atoms and the Rydberg atom, respectively, and $m_e$ is the electron mass. The Rydberg electron probability density $\rho_e(\vec r_i-\vec r_R)$ at the position of atom $i$ is determined using an analytical approximation to the radial wavefunctions of Rydberg states from quantum defect theory~\cite{Kostelecky1985} and the spherical harmonic functions for $J=1/2,m_j=1/2$ states. The Rabi dynamics are then simulated assuming perfect blockade, but where the $(N-1)$-fold excited-state degeneracy is broken by $\Delta_R$. The single-excitation efficiency $p_1(\tau)$ is then simulated numerically and averaged over 1000 random configurations for each value of the principal quantum number $n$. For $n\lesssim 30$ the probability for two atoms to overlap within the Rydberg orbital radius is vanishingly small, while for $n\gtrsim 38$ electron-atom scattering can significantly reduce the contrast of the collective Rabi oscillations.

\begin{figure}[htb]
\centering
\includegraphics[width=  00.99\columnwidth]{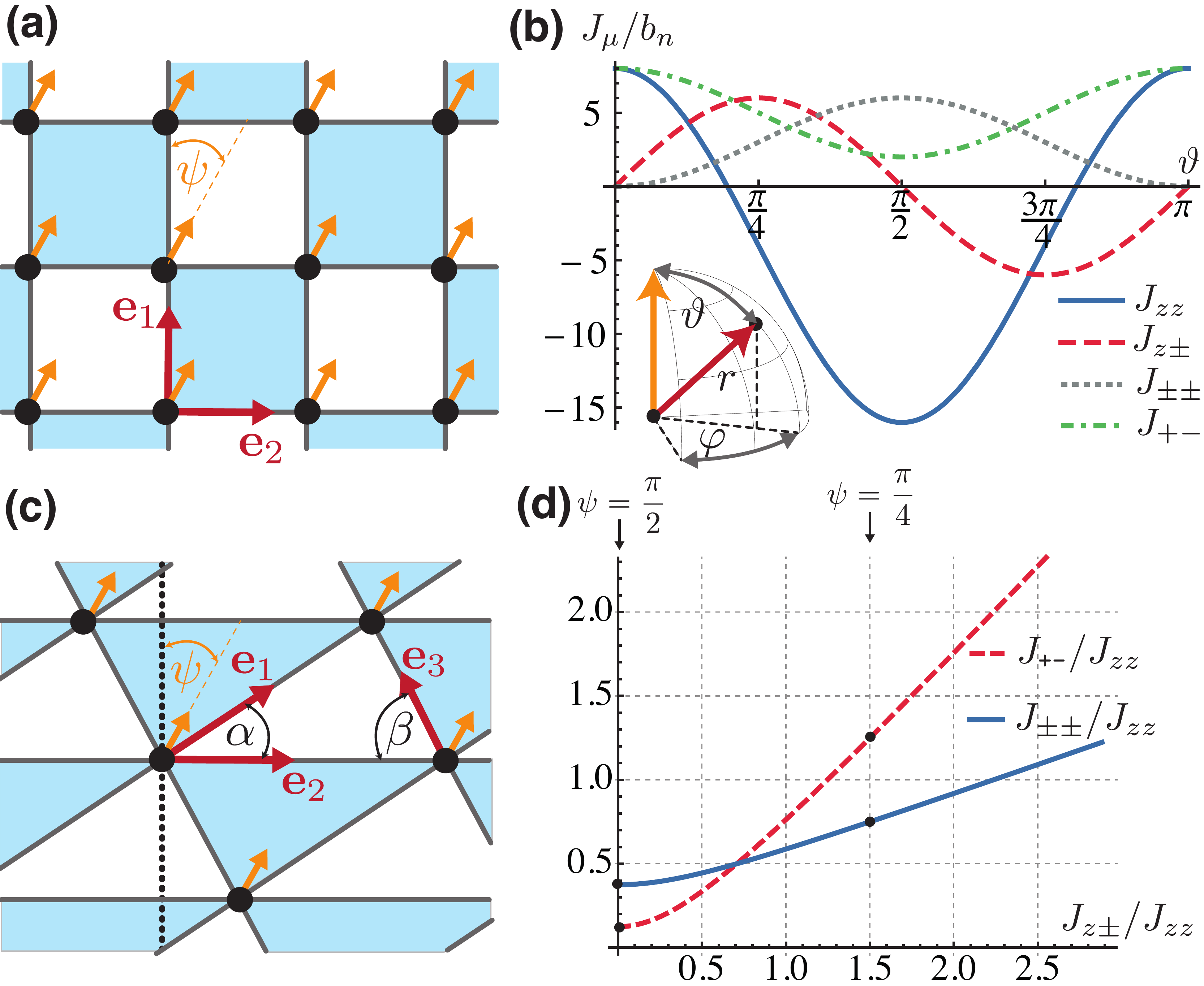}
\caption{\small{Tuning of anisotropic spin-spin interactions via the lattice symmetry and quantisation axis. (a) Spins (black dots) located on a rectangular lattice with orthogonal bond directions $\mathbf{e}_1$ and $\mathbf{e}_2$ (red arrows) and quantisation axis tilted by an angle $\psi$ (yellow arrow). (b) Interaction coefficients of Eq.~\eqref{eq:Jcoeff} as a function of $\vartheta$. (c) Spins (black dots) on a non-equilateral triangular lattice with three bond directions $\mathbf{e}_1$, $\mathbf{e}_2$ and $\mathbf{e}_3$ (red arrows) and in-plane quantisation axis tilted by an angle $\psi$ (yellow arrow). (d) Variation of the relative strength of the interaction coefficients of Eq.~\eqref{eq:Jcoeff} as a function of $\psi$ for bond 1 of the square lattice. }}
\label{fig:spinmodels}
\end{figure}
Overall these simulations indicate that for the parameters of the magnetic lattice microtraps, the optimal $n$ for maximising the efficiency of initial state preparation for $N$ between 5 and 15 atoms is around $n\approx 36$ with an estimated overall efficiency around $\langle p_1\rangle=0.87$. This clearly exceeds the classical percolation threshold (indicating the transition to long-range connectivity) for a 2D triangular lattice expected for an occupation probability of 0.5. It is also close to the state-of-the-art in single-atom preparation in optical microtraps using light-assisted collisions, where efficiencies up to 90\% have recently been achieved~\cite{Fung2016}. Furthermore, the fall-off of the efficiency with $n$ for high $n$ is slow, e.g., the efficiency at $n=60$ is still 80\%. It is {likely }that the collective state preparation efficiency can be increased to even higher values using, e.g., adiabatic state preparation or composite pulse sequences~\cite{Beterov2013}.
 
\section{Long-Range and anisotropic spin-spin Interaction}\label{sec:spin_interactions}

Given that the array sites can be initialised with high occupation probability, we now turn to the realisation of lattice spin models where the spin-1/2 degree is encoded in the collective spin states of Eq.~\eqref{eq:spins}. With Rydberg $nP_{1/2}$ states one is able to realise the most general spin-1/2 exchange Hamiltonian where spin degrees of freedom are encoded in the two Zeeman sublevels and vdW interactions give rise to spin-spin interactions of the form~\cite{Glaetzle2015,Vermersch2015}
\begin{equation}
\begin{split}
\label{eq:spinmodel}
H=\frac12&\sum_\mu\sum_{i,j\in\mu}\frac{1}{r_{ij}^6}\left[J_{zz}^{(\mu)}(\vartheta_\mu)\mathsf{S}^z_i\mathsf{S}^z_j\right.\\
-&J_{+-}^{(\mu)}(\vartheta_\mu)\left(\mathsf{S}^+_i\mathsf{S}^-_j+\mathsf{S}^-_i\mathsf{S}^+_j\right)\\
+& J_{\pm\pm}^{(\mu)}(\vartheta_\mu)\left(e^{-2i\varphi}\mathsf{S}^+_i\mathsf{S}^+_j+e^{2i\varphi}\mathsf{S}^-_i\mathsf{S}^-_j\right)\\
+& \left.J_{z\pm}^{(\mu)}(\vartheta_\mu)\left[\mathsf{S}^z_i\left(e^{-i\varphi}\mathsf{S}^+_j+e^{i\varphi}\mathsf{S}^-_j\right)+{\rm h.c.}\right]\right].
\end{split}
\end{equation}
Here, $\mathsf{S}^{z}_i$ denotes the z-component of the spin-1/2 operator and $\mathsf{S}^{\pm}_i$ denotes the spin raising/lowering operator at lattice site $i$ while $(r,\vartheta,\varphi)$ are the spherical components of the relative vector connecting spins $i$ and $j$. With $\mu$ we denote the bond of the lattice along the unit vector $\mathbf{e}_\mu$ as illustrated in Fig.~\ref{fig:spinmodels}. For a square lattice $\mu\in\{1,2\}$ while for the triangular lattice $\mu\in\{1,2,3\}$, see panels (a) and (c) of Fig.~\ref{fig:spinmodels}, respectively. The coupling constants of Eq.~\eqref{eq:spinmodel} are given by (see Appendix~\ref{app:vdW})
\begin{equation}
\begin{split}
&J_{zz}^{(\mu)}(\vartheta)=[12\cos(2\vartheta)-4]\;b_n,\\
&J_{+-}^{(\mu)}(\vartheta)=[3 \cos (2 \vartheta )+5]\;b_n,\\
&J_{\pm\pm}^{(\mu)}(\vartheta)=6  \sin ^2(\vartheta )\; b_n,\\
&J_{z\pm}^{(\mu)}(\vartheta)=6 \sin (2\vartheta ) \;b_n,\\
\label{eq:Jcoeff}
\end{split}
\end{equation}
with $b_n>0$ (for $nP_{1/2}$ states) being generalised vdW coefficients [see Eqs.~\eqref{eq:abp} and~\eqref{eq:abs} in Appendix~\ref{app:vdW}] which determine the overall sign and strength of the spin-spin interactions. The angle $\vartheta$ is the angle between the quantisation axis (yellow arrow) and the relative vector connecting the two atoms (red arrow) along the bond $\mu$, see inset in Fig.~\ref{fig:spinmodels}(b). The model Hamiltonian of Eq.~\eqref{eq:spinmodel} can serve as a toolbox for studying general spin-spin models in which the nature of the couplings can be different along different lattice directions.

As a particular example to demonstrate the tunability of the resulting spin-interaction toolbox we discuss square and triangular lattices which have already been realised with magnetic trap arrays~\cite{Leung2014}. In the conceptionally simplest case where the quantisation axis defined by the magnetic field direction at the trap bottom is aligned perpendicular to the 2D chip ($\vartheta=\pi/2$), we obtain an XXZ~Heisenberg model with $J_{zz}=-16b_n$, $J_{\pm\pm}=6b_n$, $J_{+-}=2b_n$ and $J_{z\pm}=0$ independent of the bond direction and lattice geometry. For $b_n>0$ (as is the case for $nP_{1/2}$ Rydberg states) it supports a ferromagnetic ground state which competes with the $J_{\pm\pm}$ term which tries to `melt' the ferromagnet through pair-correlated spin flips. By breaking the lattice symmetry (e.g., moving to rectangular or non-equilateral triangle lattices) one can tune the relative strength of the interaction coefficients along different bond directions due to the strong $1/r^6$ dependency. Additionally, the effects of the $J_{\pm\pm}$ terms can be suppressed by increasing the magnetic field at the trap bottoms which results in an excitation energy gap for non-spin conserving terms.

Even richer spin models can be studied by aligning the quantisation axis in-plane giving rise to anisotropic spin models for various geometries, where the interaction coefficients of Eq.~\eqref{eq:Jcoeff} vary as a function of the bond directions labelled by $\mathbf{e}_\mu$.
We first consider the square lattice of Fig.~\ref{fig:spinmodels}(a) with orthogonal bond directions $\mathbf{e}_1$ and $\mathbf{e}_2$ and an in-plane quantisation axis tilted by an angle $\psi$. The relative angles for the two bonds $\mu=1,2$ of Eq.~\eqref{eq:spinmodel} are $\vartheta_1=\psi$ and $\vartheta_2=\pi/2-\psi$. In the special case of $\psi=0$ the resulting spin-spin interactions for the first bond (i.e., $\vartheta_1=0$) correspond to an XXZ Heisenberg Hamiltonian with $J_{zz}=J_{+-}$ and $J_{\pm\pm}=J_{z\pm}=0$. However, for the second bond direction $\vartheta_2=\pi/2$, which gives rise to additional $J_{\pm\pm}$ terms (Fig.~\ref{fig:spinmodels}b). By rotating the quantisation axis such that $\psi>0$ one can increase the $J_{\pm\pm}$ and $J_{z\pm}$ coefficients at the expense of $J_{zz}$ and $J_{+-}$, thus offering the possibility to explore the rich phase structure of spin models with anisotropic couplings which are expected to exhibit a variety of non-trivial ground states~\cite{Savary2012,Savary2013}. This is illustrated for the first bond direction of the square lattice in Fig.~\ref{fig:spinmodels}(d) where by changing $\psi$ from $\pi/5$ to $\pi/2$ one can change the interaction coefficients according to Eq.~\eqref{eq:Jcoeff} over a wide range.

For the triangular lattice illustrated in Fig.~\ref{fig:spinmodels}(c) one can implement spin models where the interaction coefficients of Eq.~\eqref{eq:Jcoeff} depend on all three bond directions $\mu=1,2,3$ with $\vartheta_1=\pi/6-\psi$, $\vartheta_2=\pi/2-\psi$ and $\vartheta_3=\pi/6+\psi$. Thus, Rydberg spins in magnetic microtrap arrays may serve as the first concrete realisation of exotic spin models such as generalised compass type models~\cite{Nussinov2015}. This could be of importance for understanding non-trivial phases of frustrated magnetism in which the competition between spin-spin interactions cannot be simultaneously satisfied for all spin pairs~\cite{Lewenstein2007,Sachdev1999,Glaetzle2015,Balents2010,Glaetzle2014}. Examples of frustrated-spin quantum magnetism are quantum spin-ice, where the spins are highly correlated but fluctuate strongly before becoming ordered at low temperatures, and fluid-like quantum spin-liquids in which the fluctuating correlated spins persist down to zero temperature~\cite{Lewenstein2007,Sachdev1999,Balents2010}, as found in real materials like rare-earth pyrochlores~\cite{Savary2012,Savary2013}. 

\begin{figure}[tb]
\centering
\includegraphics[width= \columnwidth]{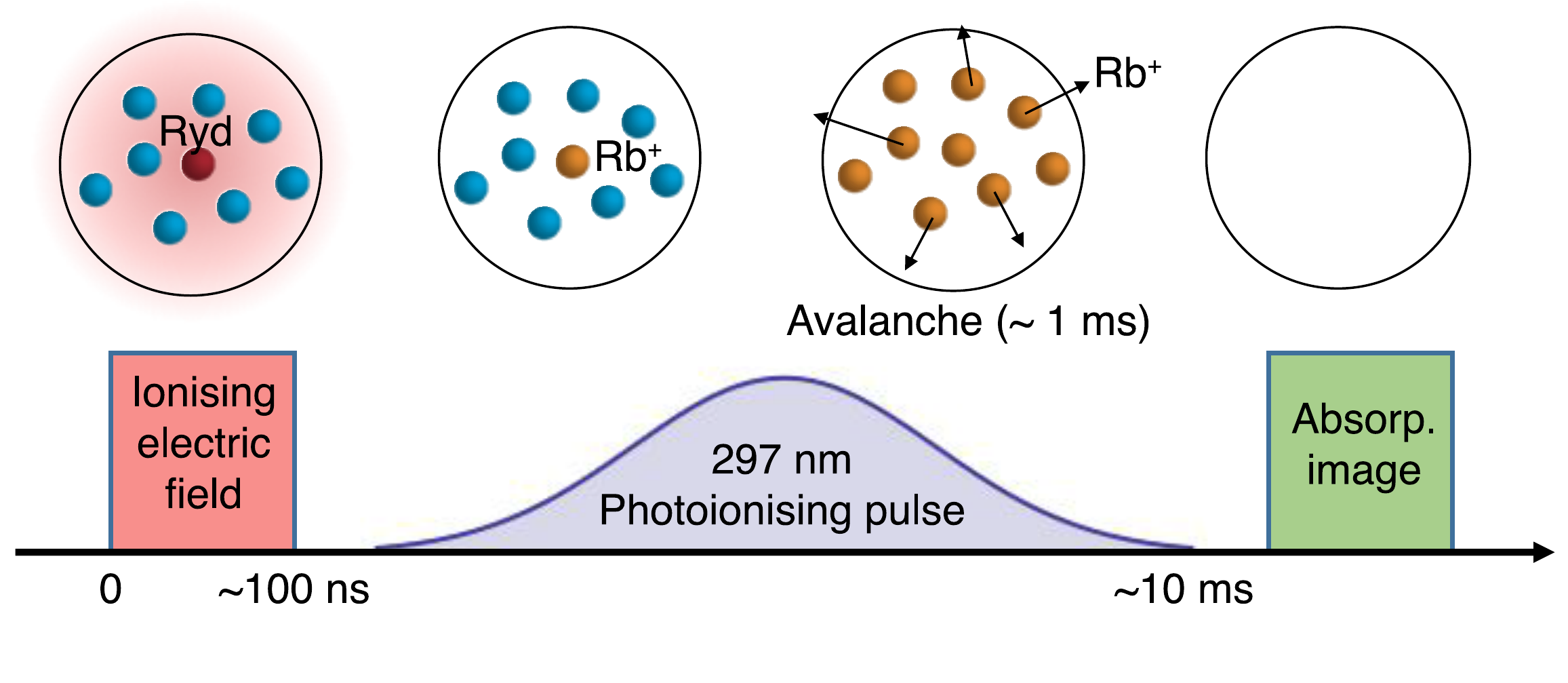}
\caption{\small{Schematic of the single Rydberg-atom triggered ionisation avalanche scheme in which the presence of a single Rydberg atom in a given spin state conditionally transfers a large number of ground-state atoms in the trap to an untrapped state via a seeded photoionisation avalanche. Site-resolved spin-spin correlations can then be directly measured via absorption imaging of the remaining filled (or empty) sites.}}
\label{fig7}
\end{figure}

\section{Readout of Spin-Spin Correlations}\label{sec:readout}
To read out the spin state in each microtrap with high fidelity we propose a triggered ionisation avalanche detection scheme which exploits each atomic ensemble as a highly sensitive amplifier. It is similar in spirit to readout schemes based on interaction enhanced imaging~\cite{gunter2012interaction,gunter2013observing} or conditional Raman transfer of the ensemble of atoms between ground states proposed in~\cite{Mueller2009}. Although the ionisation approach is inherently destructive, it may be more robust than other schemes which rely on coherent control of Rydberg states and it does not require that the participating states remain magnetically trapped. Similar ionisation avalanche processes have already been observed in experiments, and appear to be a very rapid way to empty a trap of atoms~\cite{Vincent2013}. We propose the following experimental procedure (graphically depicted in Fig.~\ref{fig7}):

\vskip 5pt
\noindent {\it 1. Spin-selective optical pumping.} First, a single spin state must be selected for detection. This is possible by coupling the $nP_{1/2},m_j=+ 1/2$ spin-up state to a lower short-lived state, such as the $6S_{1/2}$ state  (lifetime 46~ns) via a resonant laser field at 739~nm (depicted in Fig.~\ref{fig1}-inset). The $6S_{1/2}$ state spontaneously decays to the short-lived $5P$ states leaving only the spin-down state in the trap. 

\begin{figure}[tb]
\centering
\includegraphics[width= 0.8\columnwidth]{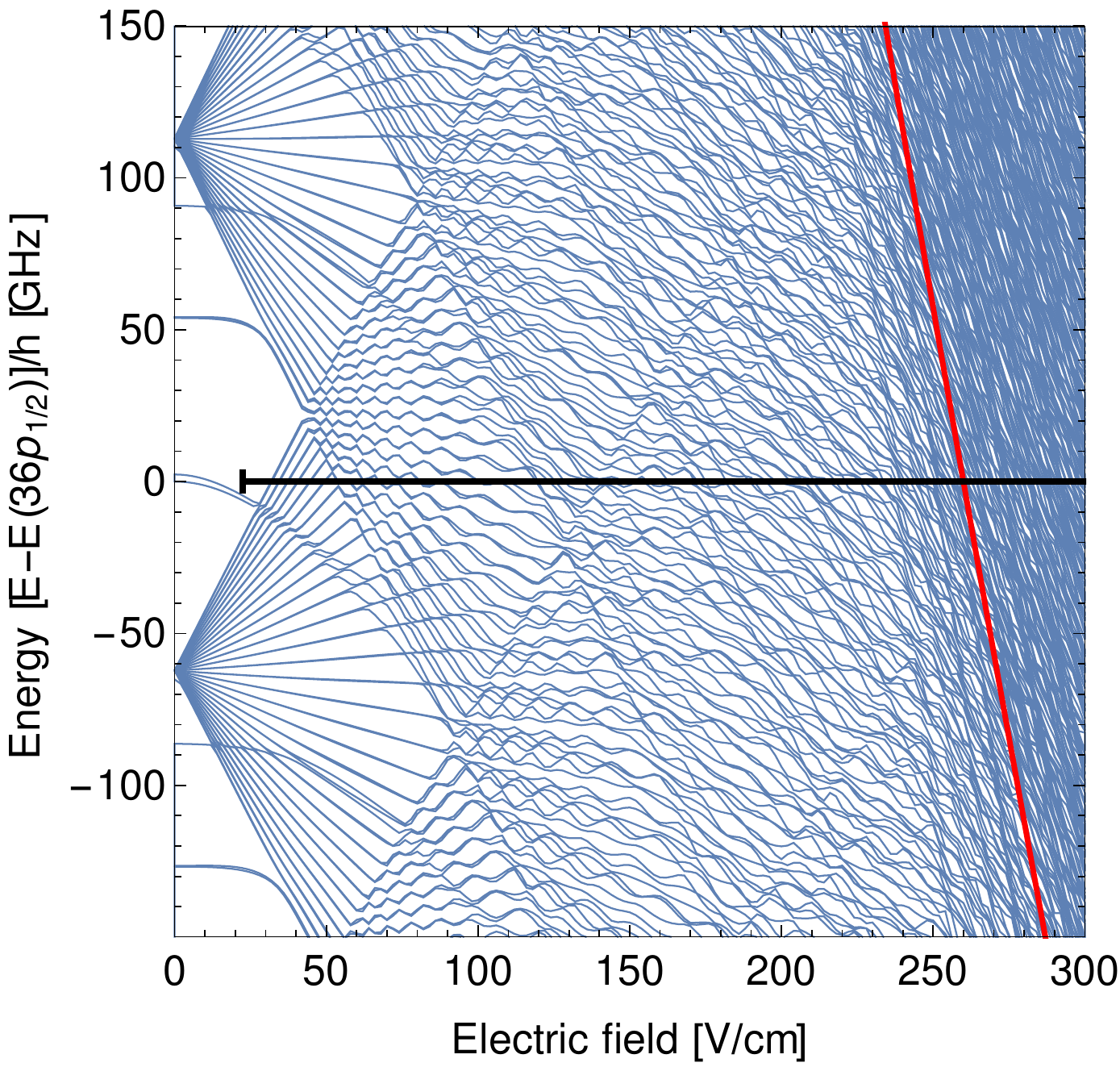}
\caption{\small{Energy as a function of electric field (Stark map) around the $36P$ state of rubidium. The red diagonal line shows the classical field-ionisation threshold. A single ion in the microtrap produces a spatially inhomogeneous electric field which decays as $1/r^2$ and ranges from $22.5\,$V/cm (corresponding to $r=l=0.8\,\mu$m) to above the classical field-ionisation threshold, to $640\,$V/cm for distances corresponding to the Wigner-Seitz radius $r=0.15\,\mu$m) as shown by the black horizontal bar. The electric field strength at neighbouring traps $F=2.3\,$V/cm is well below the classical field-ionisation threshold.}}
\label{fig8}
\end{figure}

\vskip 5pt
\noindent {\it 2. Pulsed field ionisation.} Next an electric field pulse with a short duration $\sim$100~ns is applied with a field strength of approximately $F=300\,$V/cm. This is sufficient to suddenly field-ionise the remaining Rydberg atoms (Fig.~\ref{fig8}). This pulse is long enough to eject the electron but short enough that the force on the produced Rydberg ions does not displace them far from the vicinity of their original microtraps. We estimate the escape time for an ion as several microseconds. These `seed' ions produce a spatially inhomogeneous electric field which decays with distance $r$ according to $q/(4\pi\epsilon_0 r^2)$, with $q$ the elementary charge and $\epsilon_0$ the vacuum permittivity.

\vskip 5pt
\noindent {\it 3. Seeded photoionisation avalanche.} A 297~nm laser pulse is then immediately applied to the whole lattice for several milliseconds. The spatially inhomogeneous field originating from the seed ions shifts the ionisation thresholds for the remaining ground-state atoms such that they can be photoionised with the 297~nm laser (Fig.~\ref{fig7}). This rapidly produces more ions which escape the magnetic trap until it is emptied. The classical field-ionisation limit, given by $E_c=-2\sqrt{F}$ (atomic units), for an energy corresponding to the $36P_{1/2}$ state corresponds to a field strength of $F=260\,$V/cm. At or above this field strength it should be very efficient to excite the remaining ground state atoms to unbound continuum states by turning the 297~nm Rydberg excitation laser back on. This is a convenient value which falls between the field strength for atom-ion separations given by the Wigner-Seitz radius and the trap length $l=0.8\,\mu$m ($640\,$V/cm or $25\,$V/cm, respectively). Early studies of strong field photoionisation of rubidium atoms using pulsed laser fields found rates exceeding $5\times 10^4\,\mathrm{s}^{-1}$~\cite{Freeman1979}, indicating that the entire trap could be photoionised in this way within a few milliseconds. To ensure there is no crosstalk between the traps we require that the electric field originating from one trap is sufficiently small at neighbouring traps to suppress photoionisation. This is satisfied for a $a=2.5\,\mu$m period lattice, for which the ion-produced field of $2.3\,$V/cm is 100 times smaller than the classical field ionisation limit for the $36P_{1/2}$ state (assuming a single charged particle).

\vskip 5pt
\noindent {\it 4. Absorption imaging of the remaining atoms.} To measure the magnetisation and spin-spin correlations across the magnetic lattice, standard {\it in-situ} reflection absorption imaging used for atom chips can be employed~\cite{Jose2014,Surendran2015,Whitlock2009}, in which a strongly absorbing or a non-absorbing site signifies $\mid\uparrow\rangle$ or  $\mid\downarrow\rangle$, respectively. Sensitive absorption imaging down to fewer than 10 atoms in each site of a magnetic lattice has already been demonstrated~\cite{Whitlock2009}.

In this way it will be possible to read out the $\mathsf{S}^{z}$ projection of the spin state in each site of the whole lattice. By repeating such experiments it will then be straightforward to calculate spin-spin correlations, such as the pairwise correlation function $\langle \mathsf{S}^{z}_i\mathsf{S}^{z}_j\rangle$ or even higher order correlation functions, which {is} sufficient to identify spin-liquid behaviour~\cite{Endres2011}. The time resolution, which is determined by the time required to field-ionise the Rydberg state (thereby freezing the spin dynamics), is typically an order of magnitude shorter than the expected time scale for nearest-neighbour spin exchange ($\sim1~\mu$s). To control errors due to imperfect initialisation or loss from the Rydberg state during evolution it could be advantageous to repeat the whole detection process by optical pumping the other spin state. 

\section{Additional Experimental Considerations}
\subsection{Magnetic lattices}
In a magnetic lattice, atoms oriented in low magnetic field-seeking states ($m_Fg_F>0$) are repelled by the increasing magnetic field in the traps allowing these atoms to be trapped in the magnetic field minima. The algorithm of Schmied et al.~\cite{Schmied2010} can be used to design optimised magnetic microstructures to create 2D magnetic lattices of various geometries, including those proposed in Sec.~\ref{sec:spin_interactions}. Magnetic lattices can be readily constructed with a large lattice spacing which allows atomic ensembles in individual sites to be easily resolved {\it in situ} using standard optical imaging.

For the quantum simulation of lattice spin models we envisage initially creating a $5\times5$ mm$^2$,  {$2.5\;\mu$m-period} triangular magnetic lattice using magnetic microstructures fabricated by electron beam lithography and reactive ion etching of a 20~nm-thick multi-atomic-layer Co/Pd film~\cite{Herrera2015}. These films have a large perpendicular magnetic anisotropy, high saturation magnetisation ($\sim$5.9~kG) and coercivity ($\sim$~1~kOe), a very small grain size ($\sim$~7~nm), and are capable of producing magnetic microstructures with very homogeneous magnetic potentials~\cite{Herrera2015}. We have simulated a triangular lattice based on the code of Schmied et al.~\cite{Schmied2010}. Assuming a period of $a=2.5\,\mu$m the trap symmetry and depth is optimal for a trap height $a/2$ above the magnetic microstructure. For an in-plane bias field of $B_\perp = 2.4$~G the corresponding magnetic traps are cigar-shaped with radial and axial trap frequencies $\omega_{\rm rad}/2\pi=64$~kHz and $\omega_{\rm ax}/2\pi=16$~kHz and a trap depth of 3.2~G (108 $\mu$K for $F=1$, $m_F=-1$). The Ioffe field $B_{\rm Ioffe} = 0.9$~G is oriented along the trap long axis which is tilted at an angle of $\psi=5\pi/12$. Based on our recent experience with one-dimensional magnetic lattices~\cite{Surendran2015}, we expect about $10^6$ $^{87}$Rb atoms in the $|F=1,m_F=-1\rangle$ low field-seeking state can be loaded to the central $200\times 200$ sites of the triangular magnetic lattice using a Z-wire microtrap. With a subsequent radio-frequency evaporative cooling phase we anticipate that each lattice site may be populated by approximately 10 atoms at a temperature $1\;\mu$K.

\subsection{UV laser excitation system}
Single-step excitation to the rubidium $nP$ Rydberg states requires high power, narrow-bandwidth UV radiation in the range 298 to 297 nm for $n = 30$ to $\infty$, see Fig.~\ref{fig1}. Similar excitation schemes have recently been employed to realise spin systems by groups at the University of New Mexico~\cite{Jau2016} and MPQ Munich~\cite{Zeiher2016}. A laser system consisting of a 1188~nm narrow-bandwidth master diode laser plus tapered amplifier system and two second harmonic generation ring resonators to generate $\sim$0.4 W of single-frequency 297~nm radiation is commercially available (Toptica Photonics). We estimate that 0.1~W of 297~nm radiation in a 1~mm-diameter beam can produce a Rabi frequency of $\sim$~1~MHz for the $5S_{1/2} - 60P_{1/2}$ transition which is sufficient for efficient excitation of high $nP$ Rydberg states.

\subsection{Surface effects}
A potential issue when using long-range interacting Rydberg atoms stored in a magnetic lattice is the effect of the atom chip surface on the Rydberg atoms, which are trapped at a height of typically one-half of a lattice spacing from the magnetic surface~\cite{Leung2014}. A main concern is that following each cooling and trapping sequence alkali atoms can stick to the surface of the atom chip to create inhomogeneous electric fields. The valence electron of each adsorbed atom can reside partially inside the metal surface and the charge separation creates a dipole whose strength is related to the difference between the work function of the metal ($5.1$~eV for gold) and the ionisation potential of the atoms (4.2~eV for Rb)~\cite{McGuirk2004}. The dipoles produce inhomogeneous electric fields that can perturb the nearby Rydberg atoms~\cite{Schmied2010}. Studies of Rb Rydberg atoms trapped at distances down to 20~$\mu$m from a gold-coated atom chip surface have revealed small distance-dependent energy shifts of $\sim\pm10$~MHz for $n \approx 30$~\cite{Tauschinsky2010}. Recent studies have demonstrated that the stray electric fields can be effectively screened out by depositing a thin ($\sim$90~nm) uniform film of Rb ($\varphi = 2.3$~eV) over the entire gold surface of a cryogenic atom chip~\cite{Hermann2014} or by using a smooth monocrystalline quartz surface film coated with a monolayer of Rb adsorbates~\cite{Sedlacek2016}. Another potential issue is the effect of the 297~nm UV beam when the beam is parallel to the chip surface at a height of about one-half of a lattice spacing. The 297~nm beam can eject electrons from the surface which in turn can perturb the Rydberg atoms.  Further work will be required to fully understand and control these surface effects.

\section{Summary and Outlook}
We have proposed a scheme to simulate lattice spin models based on the use of strong and long-range interacting Rydberg atoms stored in a large-spacing magnetic lattice. We point out, however, that these ideas could equally well be implemented in large period optical lattices. Each spin is encoded directly in a collective spin state involving a single $nP$ Rydberg atom in an ensemble of ground-state rubidium atoms prepared via Rydberg blockade. The Rydberg spin states on neighbouring lattice sites are allowed to interact with the driving fields turned off. Afterwards they are read out using a single-Rydberg atom triggered photoionisation avalanche scheme in which the presence of a single Rydberg atom conditionally transfers a large number of ground-state atoms in the trap to untrapped states which can be readily detected by standard site-resolved absorption imaging.

The use of Rydberg states leads to spin-spin coupling strengths which are much larger than the relevant decoherence rates and provides a way to design and realise complex spin models including Heisenberg-like spin models with isotropic and anisotropic interactions. This paves the way towards engineering exotic spin models, such as spin models based on triangular-based lattices which can give rise to a rich quantum phase structure including frustrated-spin states. Experiments could probe spin-spin correlations on different spatial scales which can be compared with theoretical descriptions to reveal the universal characteristics of these systems including ground-state properties, critical exponents and relaxation dynamics.

In addition to the ground-state phase diagram, this quantum simulator is suited to study transient many-body phenomena. Given that the lifetimes of the high Rydberg states are typically $>100\,\mu$s and the characteristic time-scale associated with spin-spin interactions is $\sim 1\;\mu$s it should be possible to investigate dynamics on both short and long time scales. In particular, it should be possible to investigate the build-up of spin-spin correlations on different length and time scales following a dynamical change in the system parameters, including their dependence on the transition rate which can be compared with, for example, the Kibble-Zurek scaling law for a system driven through a continuous phase transition at finite rate~\cite{Dutta2010}. Other interesting questions that could be addressed are how long and by what path does a far-from-equilibrium isolated quantum system take to reach an equilibrium state? and is it possible to connect certain non-equilibrium properties to the properties of the ground state?

\section{Acknowledgments}
We are indebted to Peter Zoller for helping initiate this project. We also thank Rick van Bijnen for use of his code for calculating Rydberg-Rydberg interaction potentials and Yun Li for stimulating discussions. This work is supported by an Australian Research Council Discovery Project grant (DP130101160). SW acknowledges support by the Heidelberg Center for Quantum Dynamics, the European Union H2020 FET Proactive project RySQ (grant N. 640378), the Deutsche Forschungsgemeinschaft under WH141/1-1 and the DFG Collaborative Research Centre ``SFB 1225 (ISOQUANT)''. AWG acknowledges support from the SFB FoQuS (Austrian Science Fund FWF Project No. F4016- N23) and the European Research Council Synergy Grant UQUAM.

\appendix
\section{Rydberg interaction between $^{87}$Rb atoms in $nP_{1/2}$ states}
In assessing the relevant interactions between Rydberg states we distinguish between two main regimes: (i) long-range van der Waals interactions, (ii) cross over to dipole-dipole interactions and avoided crossings and spaghetti physics.

\begin{figure*}[tb]
\centering
\includegraphics[width= 1.5\columnwidth]{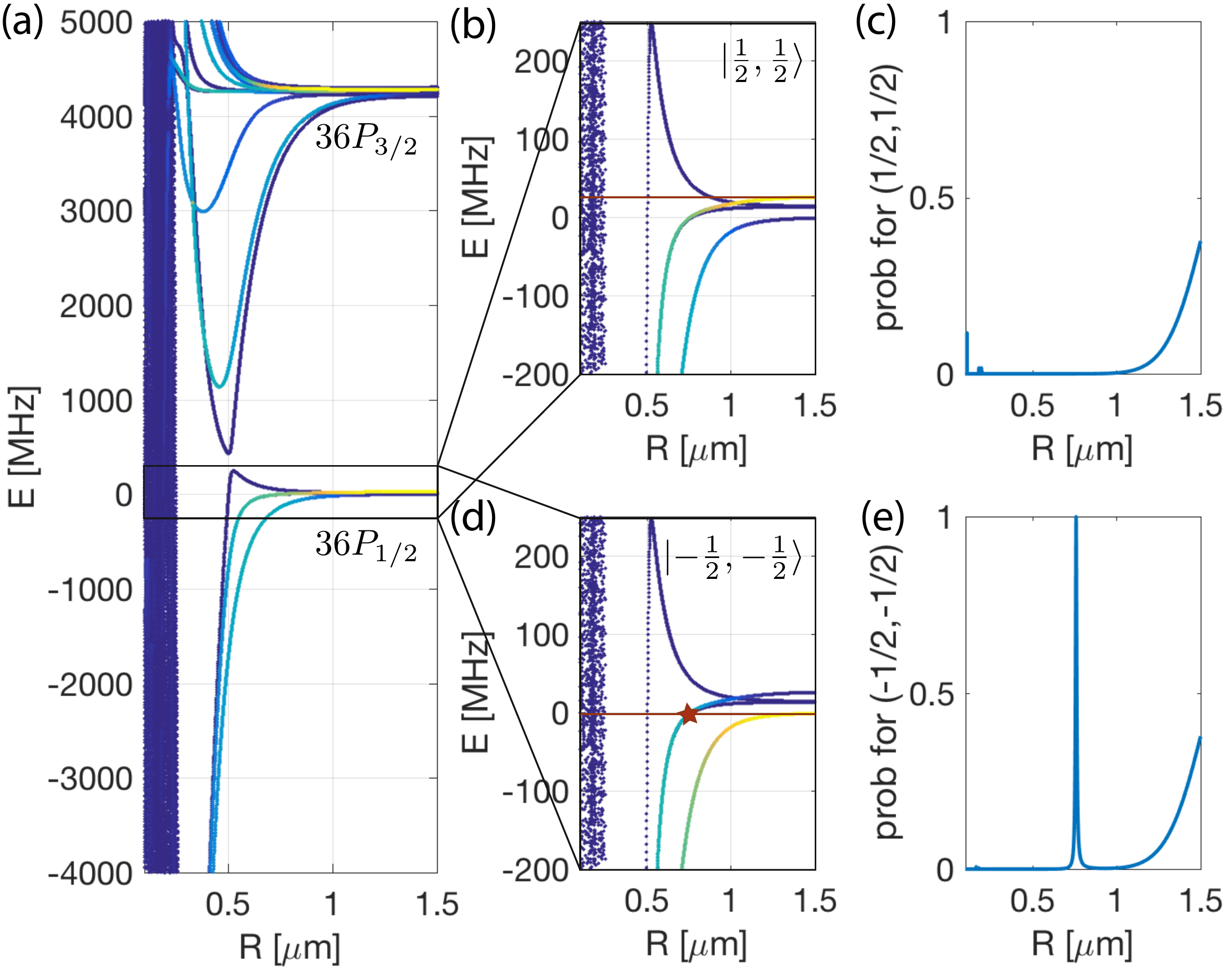}
\caption{\small{(a) Born Oppenheimer interaction potentials around the $2\times 36P_{1/2}$ Rydberg states. (b,d) Magnification of the interaction potentials. The red line corresponds to the laser excitation energy. The color quantifies the contribution of the $|\po\po\rangle$  [panel~(b)] and $|\mo\mo\rangle$ [panel~(d)] Zeeman sublevels to the Born Oppenheimer eigenstates. (c,e) Excitation probability for double excited Rydberg states as a function of distance.}}
\label{fig:spaghetti}
\end{figure*}

\subsection{Van der Waals interactions}\label{app:vdW}
The total vdW interaction Hamiltonian between two atoms in a Rydberg $nP_{1/2}$ subspace is
\begin{equation}\label{eq:Hd}
H_{\rm vdW}^{(n)}=\frac{a_n}{r^6}\mathbb{1}_4\\
+\frac{b_n}{r^6}\mathcal{D}_0,
\end{equation}
with generalised van der Waals coefficients for $nP_{1/2}$ Rydberg states of $^{87}$Rb
\begin{widetext}
\begin{equation}
\begin{split}
&a_n=\left(E_H a_0^6\right)\left[-19.0+0.50\;n- 1.14\cdot 10^{-2}\;n^{2}+1.09\cdot 10^{-4}\;n^3-3.77\cdot 10^{-7}\;n^{4}\right]n^{11},\\
&b_n=\left(E_H a_0^6\right)\left[0.93- 1.60\cdot 10^{-2}\;n+4.88\cdot 10^{-4}\;n^2-5.08\cdot 10^{-6}\;n^{3}+1.83\cdot 10^{-8}\;n^4\right]n^{11},
\label{eq:abp}
\end{split}
\end{equation}
and for $nS_{1/2}$ Rydberg states 
\begin{equation}
\begin{split}
&a_n=\left(E_H a_0^6\right)\left[-13.8.0+0.92\;n- 3.45\cdot 10^{-3}\;n^{2}-1.34\cdot 10^{-5}\;n^3+1.08\cdot 10^{-7}\;n^{4}\right]n^{11},\\
&b_n=\left(E_H a_0^6\right)\left[8.19\cdot 10^{-3}+ 4.45\cdot 10^{-4}\;n-5.72\cdot 10^{-5}\;n^2+5.72\cdot 10^{-7}\;n^{3}-1.89\cdot 10^{-9}\;n^4\right]n^{11},
\label{eq:abs}
\end{split}
\end{equation}
which strongly depend on the principal quantum number $n$. Here, $E_H$ is the Hartree energy and $a_0$ is Bohr's radius, i.e., $E_H a_0^6/2\pi=1.44\times 10^{-16}$~MHz~$\mu$m$^6$. With $\mathbb{1}_4$ we denote the $4\times 4$ identity matrix and 
\begin{equation}
\mathcal{D}_0(\vartheta,\varphi)=
\left(
\begin{array}{cccc}
 3 \cos (2 \theta )+11 & 3 e^{-i \phi } \sin (2 \theta ) & 3 e^{i \phi } \sin (2 \theta ) & 6 e^{-2 i \phi } \sin ^2(\theta ) \\
 3 e^{i \phi } \sin (2 \theta ) & 13-3 \cos (2 \theta ) & -3 \cos (2 \theta )-5 & -3 e^{i \phi } \sin (2 \theta ) \\
 3 e^{-i \phi } \sin (2 \theta ) & -3 \cos (2 \theta )-5 & 13-3 \cos (2 \theta ) & -3 e^{-i \phi } \sin (2 \theta ) \\
 6 e^{2 i \phi } \sin ^2(\theta ) & -3 e^{-i \phi } \sin (2 \theta ) & -3 e^{i \phi } \sin (2 \theta ) & 3 \cos (2 \theta )+11 \\
\end{array}
\right)
\label{eq:D0}
\end{equation}
\end{widetext}
written in the basis $\{|\po \po\rangle,|\po\mo\rangle,|\mo\po\rangle,|\mo\mo\rangle\}$ of Zeeman states accounts for the anisotropy and Zeeman mixing of the van der Waals interactions. Within an arbitrary energy offset Eqs.~\ref{eq:Hd} and \ref{eq:D0} are equivalent to the spin-spin interaction Hamiltonian in the main text (Eq.~\ref{eq:spinmodel}). Diagonalising $H_{\rm vdW}^{(n)}$ yields the isotropic eigenenergies
\begin{equation}
\begin{split}
E_1=&E_2=(a_n+14 b_n)/r^6,\\
&E_3=(a_n+2 b_n)/r^6,\\
&E_4=(a_n+18 b_n)/r^6,\\
\end{split}
\end{equation}
with corresponding eigenstates
\begin{equation}
\begin{split}
&|E_1\rangle=\frac{1}{\sqrt{2}}\left(e^{-i\phi}|\po \po\rangle+e^{i\phi}|\mo \mo\rangle\right),\\
&|E_2\rangle= \frac{\cos\vartheta}{\sqrt{2}}\left(e^{-i\phi}|\po \po\rangle-e^{i\phi}|\mo \mo\rangle\right)+\frac{\sin\vartheta}{\sqrt{2}}\left(|\po \mo\rangle+|\mo \po\rangle\right),\\
&|E_3\rangle=\frac{\sin\vartheta}{\sqrt{2}}\left(e^{i\phi}|\mo \mo\rangle-e^{-i\phi}|\po \po\rangle\right)+\frac{\cos{\vartheta}}{\sqrt{2}}\left(|\po \mo\rangle+|\mo \po\rangle\right),\\
&|E_4\rangle=\frac{1}{\sqrt{2}}\left(|\po \mo\rangle-|\mo \po\rangle\right).\\
\end{split}
\end{equation}

\subsection{Short-range physics}\label{app:shortrange}
The perturbative treatment giving rise to vdW interactions becomes increasingly inaccurate for small interatomic distances, e.g., for $d<1\;\mu$m and $n\approx 30$. In order to obtain the interaction potentials in the regime of small interatomic distances, $d<1\;\mu$m, but still large enough such that the Rydberg orbits of size $R_{\rm ryd}\sim a_0 n^2$ do not overlap, $R_{\rm ryd}\ll d$, we diagonalise the dipole-dipole interaction Hamiltonian using $10^4$ basis states. Figure~\ref{fig:spaghetti}(a) shows a typical example of interaction potentials around the $2\times 36 P$ Rydberg states of $^{87}$Rb for $\vartheta=\pi/2$ and a magnetic field splitting of 20~MHz (14.3 Gauss). Panels (b,d) show a magnification of the interaction potentials around the $2\times 36P_{1/2}$ Rydberg state. The colour code indicates the overlap of the Born-Oppenheimer eigenstates with the $|\po\po\rangle=|36P_{1/2},m_j=1/2,36P_{1/2},m_j=1/2\rangle$ state [panel (b)] or with the $|\mo\mo\rangle=|36P_{1/2},m_j=-1/2,36P_{1/2},m_j=-1/2\rangle$ state [panel (d)]. For distances smaller than 1~$\mu$m the states $|\mo\mo\rangle$ and $|\po\po\rangle$ start to mix. Panels~(c,e) show the excitation probability of a laser resonant with the $|\po\po\rangle$ [panel (c)] or the $|\mo\mo\rangle$ [panel (e)] states as a function of the interatomic separation. For large separation and vanishing interaction energy both probabilities approach unity, while for small interatomic separations the excitation probability should vanish due to the Rydberg blockade effect. However, it could happen that at short distances accidental resonances lead to resonant pair excitations. Taking into account $10^4$ Rydberg states, panels (c,e) show that the excitation probability due to the very small overlap of the wavefunctions is negligible even for distances as large as 1~$\mu$m. In panel~(e) we observe ``magic distances''~\cite{Vermersch2015} at around $0.68~\mu$m [red star in panel (d)]. This resonance occurs due to the fact that one couples to Zeeman states with the lowest energy and thus resonantly hits the (attractive) Born-Oppenheimer potential which asymptotically connects to the $|\po\po\rangle$ Rydberg states but at short distances contains a significant contribution of the $|\mo\mo\rangle$ Rydberg state. These resonances can be avoided by using the right combination of laser polarisation and magnetic field direction.

\bibstyle{natbib}
\bibliography{spinarray}

\end{document}